# Ultra-narrow Bright Spatial Solitons Interacting with Left-handed Surfaces

**Allan Dawson Boardman, Larry Velasco, Neil King and Yuriy Rapaport**

*Joule Physics Laboratory, Institute for Materials Research*

*University of Salford, Salford, M5 4WT, UK*

A vectorial FDTD method is used to present a numerical study of very narrow spatial solitons interacting with the surface of what has become known as a left-handed medium. After a comprehensive discussion of the background and the family of surface modes to be expected on a left-handed material, bounded by dispersion-free right-handed material, it is demonstrated that robust outcomes of the FDTD approach yield dramatic confirmation of these waves. The FDTD results show how the linear and nonlinear surface modes are created and can be tracked in time as they develop. It is shown how they can move backwards, or forwards, depending either upon a critical value of the local nonlinear conditions at the interface, or the ambient linear conditions. Several examples are given to demonstrate the power and versatility of the method, and the sensitivity to the launching conditions.

## 1. INTRODUCTION

The launching of waves onto surfaces is an important problem [1,2]. In previous decades it has come under various headings but it is clear that the input-output coupling mechanism to a surface or thin film is rather critical. Classical examples [3] embrace prism, end-fire, and grating coupling. Even so, although the latter two are valuable techniques they will not be addressed in this paper, leaving the focus upon the importance of prism coupling into structures that are deemed to be particularly interesting. The emphasis here is upon the interaction of narrowly confined nonlinear light beams [4-9] with the surfaces of what has now become known as a left-handed material. It will be demonstrated that such ultra-narrow beams, called spatial solitons [10], are highly desirable for interrogating the kind of surfaces most likely to be part of a future, highly confined, all-optical chip environment. These beams strongly resist diffraction through the agency of self-focusing and they can be deployed at practical angles of incidence to the interfaces. There is no doubt that linear beams can be used to excite surface waves [1] but the use of narrow beams is clearly a practical advantage in any device application. Unfortunately an extremely narrow linear beam has a short Rayleigh length, typically the order of microns, so massive diffraction would ensue. Combating diffraction with nonlinearity is a very good way forward and some authors have referred to beams with such properties as 'optical needles' [4,5]. It is interesting that for the Kerr medium considered here that there is no lower limit of bright TE solitons but there is a limit of half a wavelength for bright TM solitons [8]. Such beams, with widths the order of a wavelength in diameter, have a behaviour that is well outside the paraxial regime. This makes them capable of being fired at surfaces with angles that are well away from the grazing incidence implied by any reliance upon the familiar nonlinear Schrödinger equation.

Left-handed materials (LHM) [11-22] are mostly discussed in terms of microwave behaviour but the search for optical frequency operation continues, especially in the area of nanophotonic structures [23-25]. In anticipation of these developments, the results given in this paper are intended for the optical regime but the general properties exposed can be scaled to apply to other regimes as well; it is, in fact, a general feature of the FDTD method that both the temporal and spatial dimensions are normalised to a chosen wavelength [26-33]. In that sense, the conclusions have a generic character for a left-handed substance that is electromagnetically linear. The left-handedness flows naturally from the assumption that the linear material *simultaneously* [11] possesses a relative permittivity $\varepsilon(\omega) < 0$ and a relative permeability $\mu(\omega) < 0$, for an operating angular frequency $\omega$. If these conditions apply, they are a *sufficient* condition for the possible excitation of *backward waves* but the inclusion of damping soon shows that these are not necessary conditions [20]. In the model adopted here, however, damping is neglected on the grounds that it can apply to many regimes of application and, more pertinently, because the qualitative overall features exposed here will be sustainable even in the presence of damping. The fact that both the permittivity and the permeability are frequency dependent makes the left-handed material dispersive and, in the optical regime, it is plasmonic [1] in character. Indeed, the dispersion is enough to require care to be exercised concerning the speed with which the beams are 'switched on' [32]. This is because a sudden introduction of a beam, to achieve the interrogation of a surface, can be accompanied by frequencies other than the main carrier frequency. Such a 'broadband' tool may impact unfavourably upon the way the beam is addressing the dispersive material. The generation of backward waves follows from the simultaneous conditions $\varepsilon(\omega) < 0$ and $\mu(\omega) < 0$, in an isotropic material, as an elementary inspection of the Maxwell curl equations for plane waves, with a wave vector $\boldsymbol{k}$, rapidly reveals. The fact that the triad ($\boldsymbol{E}, \boldsymbol{H}, \boldsymbol{k}$) is then *left-handed* is the reason why such media have come to be known as left-handed [11]. Furthermore, an electromagnetic beam upon entering a left-handed medium engages in a form of negative refraction that causes it to bend beyond the normal to the surface. This phenomenon is widely attributed to modern times but in fact was shown unequivocally by Schuster in 1904, when he was Professor of Physics at the University of Manchester [34, 35]. His conclusions were the outcome of conversations with his mathematical colleague Lamb who was then investigating the group velocity of water waves. Schuster realised, immediately, that the deviation of a wave entering a medium that supports backward waves would be greater than the angle of incidence i.e. the waves will appear to bend past the normal and be *negatively refracted*. Schuster cited absorption bands as hosts for this kind of behaviour, which may explain why the idea of using backward waves was not taken up more rapidly.

The fact that a negative, frequency-dependent $\varepsilon(\omega)$, is part of the heartbeat of using a left-handed material leads to the expectation that any surface waves that becomes established, or are still in the dynamical process of being created, when an electromagnetic beam addresses a surface ought to have something in common with the well-known linear surface plasmon-polaritons [1]. The fact that the latter are TM-polarised immediately raises a question about whether left-handed materials can also support only TM-polarised modes. The answer is that *both* TM and TE *linear* surface modes [17] may exist upon the type of left-handed material defined here because the condition $\mu(\omega) < 0$ introduces a discontinuity in the slope of the TE electric field component at the boundary, thus giving it a similar appearance to the TM mode. This will be explained later on but it is a property that will permit the use of TE-polarised beams throughout the simulations.

## 2. SURFACE WAVES

Plane interfaces, in the absence of an externally applied electric or magnetic field, are members of a class of open waveguides that can, in principle, support either transverse electric (TE) or transverse magnetic (TM) modes, but not both together [1]. The *linear case* consisting of a single interface between a semi-infinite upper bounding medium and a semi-infinite lower substrate medium has received a lot of attention over the years because it is an arrangement that supports TM-polarised surface plasmon-polaritons. A typical open-guiding system consists of a surface separating a semi-infinite, non-magnetic, dispersion-free dielectric, with relative permittivity $\varepsilon_b$, from a semi-infinite dispersive dielectric medium that has a relative permittivity $\varepsilon(\omega) < 0$, where $\omega$ is an angular frequency. Suppose that the media are of infinite extent in the $\pm y$ directions, all the field components are independent of $z$ and the surface waves propagate along the $x$-axis with a wave number $k$. A right-handed (TM) *forward surface wave*, for which $\mu(\omega) = 1$, satisfies the conditions

$$-\frac{\varepsilon(\omega)}{\varepsilon_b} > 0 ; k^2 = \frac{\omega^2}{c^2} \left\{ \frac{\varepsilon_b \varepsilon(\omega)}{\varepsilon_b + \varepsilon(\omega)} \right\} \tag{1}$$

This formula applies to what can be termed a right-handed medium (RHM) for which the relative permittivity and relative permeability are not simultaneously negative. Equation (1) is an entirely linear result emphasising that the modes *must be*, in this case, TM-polarised and that $\varepsilon(\omega)$ and $\varepsilon_b$ must be *opposite* in sign. To be locked onto the interface it is necessary that $\varepsilon(\omega) + \varepsilon_b < 0$. Hence, if $\varepsilon_b > 0$ the existence of TM surface modes is a matter of whether a material with $\varepsilon(\omega) < 0$ can be created. A metal is just such a material so this is an easy condition to create experimentally. For TE-polarised waves, however, the condition for surface wave existence is

$$\sqrt{k^2 - \frac{\omega^2}{c^2}\varepsilon_b} + \sqrt{k^2 - \frac{\omega^2}{c^2}\varepsilon(\omega)} = 0 \qquad (2)$$

but, unfortunately, for right-handed media (RHM) equation (2) has no physical solutions.

If the condition on the relative permeability is now relaxed to permit dispersion through the adoption of the *simultaneous* conditions $\varepsilon(\omega) < 0$ and $\mu(\omega) < 0$, for one of the semi-infinite media, then $(\boldsymbol{k}, \boldsymbol{E}(\omega), \boldsymbol{H}(\omega))$ is a *left-handed* set in that medium. This kind of metamaterial can support *backward* waves, for which the phase and group velocity are of opposite sign. Any surface waves can be either TM *or* TE polarised because $\mu(\omega) < 0$. This points the way to the adoption of TE-polarised spatial solitons, which introduces some ease of the computational cost when trying to demonstrate the behaviour of ultra-narrow, non-paraxial solitons in the neighbourhood of surfaces.

A possible theoretical model for a left-handed metamaterial (LHM) is the Drude form

$$\varepsilon(\omega) = 1 - \frac{\omega_{pe}^2}{\omega(\omega + i\Gamma_e)}, \quad \mu(\omega) = 1 - \frac{\omega_{pm}^2}{\omega(\omega + i\Gamma_m)} \qquad (3)$$

where $\omega$ is the angular frequency, $\omega_{pe}$ and $\omega_{pm}$ are the respective plasma frequencies and $\Gamma_{e/m}$ are damping factors. A typical value $\omega = \omega_c = 1.77 \times 10^{15}$ s$^{-1}$ is used throughout the simulations. This corresponds to a free-space wavelength of 1.064μm. The values of $\varepsilon$ and $\mu$ differ among the simulations presented and correspond to selecting metamaterials with different respective plasma frequencies. The latter can easily be obtained by working backwards from the given values using equation (3). If $\Gamma_{e/m}$ are really significant they can play a non-trivial role. This is true to such an extent that a major consequence of $\Gamma_{e/m} \neq 0$ is that a negative permeability is then not a necessary condition for the existence of backward waves [20]. In spite of this, many systems have a small enough $\Gamma_{e/m}$ for the results here to aim at physical situations for which, effectively, $\Gamma_{e/m} = 0$. It must be acknowledged, however, that for backward waves, in highly absorbing regions, of the kind highlighted by Schuster [35], damping may have qualitative as well as quantitative consequences.

Consider first a single interface between a right-handed, *linear*, non-metallic medium (RHM) for which $\mu(\omega) = 1$, $\varepsilon(\omega) > 0$. This surface *cannot* support a *linear* TE-polarised surface wave, or indeed a *linear* TM mode. The sketch shown in Fig. 1(a) has an interface between two RHM media one of which is *nonlinear*, however. The field in the RHM can then be distorted into a *sech*-type curve and the simultaneous matching of the electric field and its derivative is possible. This behaviour takes place because the nonlinearity changes the index by an amount that is proportional to the intensity of the wave. Unfortunately, there is a major downside to generating such a surface wave. A substantial minimum power level is required for its existence. It is so high, in fact, that for accessible nonlinear coefficients it is impracticable for any conceivable integrated optical device. It may not be out of reach for the noble gases, however [5].

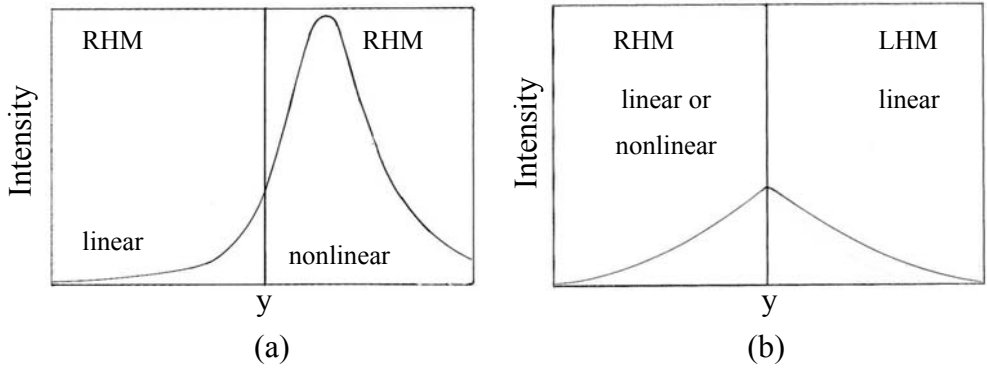

**Figure 1:** Sketches of the field profile intensities as a function of the y-coordinate normal to the boundary between (a) a RHM linear dielectric and a RHM Kerr-type nonlinear dielectric and (b)TE surface waves at a RHM linear dielectric and a LHM linear dispersive plasma). Both (a) and (b) satisfy the TE boundary conditions.

Fig. 1(b) is a sketch of TE field profiles for the LHM case. The presence of the LHM does not imply that the surface waves will always be backward travelling waves, however. This is determined by the relationship between the permittivity of the RHM and the permittivity and permeability of the LHM.

To begin this discussion, it should be noted that surface waves are evanescent and the decay is $e^{-\kappa_b y}$ for $y > 0$, $e^{\kappa y}$ for $y < 0$ where

$$\kappa_b = \sqrt{k^2 - \varepsilon_b \frac{\omega^2}{c^2}}, \quad \kappa = \sqrt{k^2 - \varepsilon(\omega)\mu(\omega)\frac{\omega^2}{c^2}} \qquad (4)$$

The second decay coefficient is for the LHM material. An important point can now be made with respect to the group velocity of the TE surface wave. Not only can this quantity be used to distinguish between forward and backward waves but also it can be manipulated to change the speed of any guided waves that are stimulated.

The dispersion equation for the TE surface waves is

$$D(\omega,k) = \kappa_b + \frac{\kappa}{\mu} = 0 \tag{5}$$

If the Drude model of the LHM is used then equation (3), without losses, is the appropriate model.

Differentiating $D$ with respect to $\omega$ and $k$ gives

$$\frac{\partial D}{\partial \omega}\delta\omega + \frac{\partial D}{\partial k_x}\delta k_x = 0 \Rightarrow v_g = \frac{\partial \omega}{\partial k_x} = -\frac{\partial D/\partial k_x}{\partial D/\partial \omega} \tag{6}$$

Hence

$$v_g = \frac{k\left(\dfrac{1}{\kappa_b} + \dfrac{1}{\mu\kappa}\right)}{\dfrac{\omega}{c^2}\left(\dfrac{\varepsilon_b}{\kappa_b} + \dfrac{1}{\mu\kappa}\right) - \dfrac{\omega_{pm}^2}{\omega^3\mu\kappa c^2}\left[\omega_{pe}^2 - 2\dfrac{\kappa c^2}{\mu}\right]} \tag{7}$$

The group velocity goes to zero and actually changes sign whenever the bounding dielectric medium has a relative permittivity $|\varepsilon_b| = |\varepsilon_c| = |\varepsilon||\mu|$. At this point, $\kappa_b \to \kappa$ and hence $\mu \to -1$, which means that the crossover condition from negative to positive group velocity is simply $|\varepsilon_c| = |\varepsilon|$. Quite apart from this crossover condition equation (7) shows the sensitivity of $v_g$ to $\varepsilon_b$. This means that controlling $\varepsilon_b$ with power can change the group velocity in a critical way that should be visible in full-scale simulations. If the group velocity is going to be changed it all depends on the starting value of $\varepsilon_b$. A similar discussion looking explicitly at the dispersion curves has been given previously [15]. The question as to whether the group velocity changes sign in a nonlinear situation will be addressed later on when detailed simulations will be presented. For $\varepsilon_b < \varepsilon_c$, the surface waves will be backward travelling, while for $\varepsilon_b > \varepsilon_c$ the surface waves will be forward travelling.

## 3. LAUNCHING AND TRACKING NARROW SOLITONS

This paper investigates spatial solitons that are sufficiently narrow [4-9] in terms of their beam width ratio to the wavelength for the nonlinear Schrödinger (NLSE) equation to be completely inapplicable. The NLSE has been a popular tool [36-38] for several decades now, not least because it is not very costly in terms of computer time. Many predictions have been made on the basis of this equation but great care needs to be exercised with it. This is because, although the split-step beam propagation method looks very attractive, it is clear from the approximations leading up to the NLSE deductions that it cannot be applied in all real situations. A major drawback is that it is based upon a slowly-varying amplitude assumption so it is only valid for very small beam orientations to the propagation axis. Indeed, these angles are so small that the beam must always be essentially parallel to the propagation axis: hence the term beam propagation. In practice, some small angles, or inclinations, are tolerable but with a beam width scale in microns and a propagation scale in millimetre Rayleigh lengths it is difficult to improve in any straightforward way. As the spatial beam narrows it becomes evident that the diffraction is not taken into account properly within the NLSE formulation. In addition to this criticism, it soon becomes clear from the field equation that the derivation also neglects the divergence of the electric field, and this step leaves out an additional nonlinearly-induced diffraction [39] that opposes beam collapse. These problems have driven several investigations into what has been termed sub-wavelength spatial solitons [6-9]. This terminology is selected to emphasise the fact that the beam widths that are less than, or the order of a wavelength, are being considered. The reason for this is that such beam widths take the investigations well into the nonparaxial regime [4,5]. Even if this is done there are still approximations, such as the neglect of the divergence of the nonlinear polarisation, and some nonparaxial methods cannot handle really small beam widths. Several investigations lead to the conclusion that bright TM spatial solitons cannot be smaller in width than $\lambda/4$, where $\lambda$ is the wavelength of light, but that bright TE can, in principle, be arbitrarily narrow provided that it is accepted that such ultra-narrow TE solitons can easily be driven into an unstable state.

In the direct simulations reported here, these dynamic possibilities are borne in mind but it is also necessary to consider what difference it will make if nonlinear saturation is omitted from the model calculations. Fundamentally, the presence of saturation limits the achievable width/wavelength ratio [5], as revealed during studies of 'optical needles'. Basically, really narrow beams that reduce in width to very small fractions of a wavelength can only be achieved if the saturation intensity is tending towards infinity, or at least is of the magnitude that has come to be expected for materials like the noble gases. For these reasons, although the simulations below use TE-polarised beams in an unsaturated medium, the beam widths are kept to the order of a wavelength. The latter choice anticipates that the results should be relevant to media that have realistic saturation intensity levels.

The finite-difference-time-domain (FDTD) method is a very valuable and direct approach to the study of beam interaction with surfaces [21], so this is the computational method of choice here. It is a powerful technique that operates in the time-domain, and although it has a long history going all the way back to 1966 [30], it is a fairly recent player on the optical stage. One of the immediate problems in using this method is that it is very costly in terms of computer time but this is less of a drawback now with the current availability of modern processors. Over the last thirteen years the FDTD technique has been used to study optical pulses and beams propagating in Kerr-type materials [28, 31, 40-43]. Furthermore, in the papers [28, 31] the FDTD has revealed some new soliton behaviour that cannot be revealed using the nonlinear Schrödinger equation. An impedance-matched RHM-LHM interface has been elegantly addressed using FDTD [21]. However, this paper addresses the use of the FDTD method to simulate spatial solitons interacting with RHM-LHM surfaces. It is possible to apply FDTD to model nonlinear materials displaying an instantaneous response. To be effective FDTD must operate in this case with a small grid-step. It also appears the research community is still using the famous algorithm created by Yee [30]. The step lengths are in microns for the problems considered here and the creation of the soliton beams can be followed from birth, to such an extent that the switch-on time of the beam and its impact on any reflection, or surface wave launching outcomes can be followed in detail. The advantage of FDTD is that it is free of assumptions, it is fully capable of dealing with the frequency dispersion presented by the left-handed materials, the light beams can be ultra-narrow, and the frequency bandwidth can be very small or be broad enough to contain a frequency spread created by rapid beam switching. The FDTD deals with interface problems in a very natural way, instead of requiring a perturbative scheme used to discuss spatial soliton optical switches [44].

TE polarized beam interactions are investigated necessitating only the three field components $E_Z$, $H_X$ and $H_Y$ simulated over an XY-plane bounded by a standard perfectly matched layer (PML) absorbing boundary, optimised in width for use with nonlinear beams. A standard Cartesian square mesh was employed and measured in fractions of the wavelength, i.e, $\delta X = \delta Y = \frac{\lambda}{N_\lambda}$, where $\lambda = 1.064 \mu m$ and $N_\lambda = 33$. $N_\lambda$ is the number of sample points defining the free-space operational wavelength. It must be noted that when continuous fields are translated into discrete domains care must be taken regarding restrictions imposed by sampling theory. In order to ensure stability and convergence of the simulation, it is generally accepted that a minimum of 10 to 20 points must be used to define the free-space operational wavelength [26, 27]. Inside a medium, where the wavelength is smaller than that of free space, it is essential to ensure that the sampling criterion is still satisfied. If the medium in question is also nonlinear, the effective refractive index and hence resulting wavelength is a function of the amplitude of the field. It is possible to derive an expression to obtain the maximum allowed amplitude in the nonlinear medium, assuming a simple self-focusing nonlinearity, below which the sampling criterion can still be met. This maximum amplitude is

$$E_{MAX} = \sqrt{\frac{\frac{N_\lambda}{N_0} - n_0}{n_{2E}}} \qquad (8)$$

where $n_0$ is the linear refractive index and $n_{2E}$ is the usual nonlinear coefficient. $N_0$ is the accepted minimum number of free space sampling points. This investigation is designed to deal with materials exhibiting Kerr nonlinearity. For example, the weakly nonlinear silicate group of materials exhibit the following typical data: $n = n_o + n_{2E}E^2$, where $n_o = 1.53$ and $n_{2E} = 1.71 \times 10^{-22}$ m$^2$/V$^2$, measured at a wavelength of 1.064μm. Given these values, and assuming the more stringent $N_0$ value of 20, the maximum permitted amplitude is found to be $E_{MAX} = 2.65 \times 10^{10}$ V/m. It is clear that if, at any point during an FDTD simulation, this value of field is exceeded then stability will be at risk. Having now established the criterion for spatial stability, it is important to consider the permitted temporal spacing that will be evident during the dynamic stages of the simulation. It is known [26], that a Courant number of 0.5 is sufficient for temporal stability in a 2D FDTD simulation. This leads to a temporal spacing of $\delta T = 65$ attoseconds, where $T$ is the time coordinate measuring the beam development from the switch-on time at $T = 0$. As will be seen in the simulations some surface waves will be generated. It is obvious that attenuation in the $y$ directions, normal to the propagation direction, of the field components could reduce the intensity distribution to a sub-wavelength variation. It is important therefore to make sure that accuracy is maintained. As stated above, a spatial resolution of $\lambda/33$ is adopted here for the results displayed. Much higher resolutions up to and beyond $\lambda/100$ does not change the nature of the results.

Having discussed the stability, some questions can be asked about the possibility of using 'optical needles' or ultra-narrow spatial soliton beams in a practical situation. To do this requires materials that are sufficiently nonlinear to cope with the formation of such narrow beams and to withstand the possibility of optical breakdown. In this paper a Kerr model is adopted for computational convenience, since the introduction of saturation produces numerical complications that have not been widely discussed in the literature. Nevertheless, the question of saturation of the nonlinearity has been addressed in the context of a *theory* of 'optical needles' [5] and the dependence of the spatial soliton width and the maximum soliton intensity on beam power have been calculated. The conclusions are that saturation does limit the beam width to be the order of the wavelength but the maximum soliton intensity is not more than a factor of 2 different in the saturated and unsaturated cases. The results given here from the Kerr model are not qualitatively different from the saturated case but the maximum field intensity is high and would need a material that has a high optical damage threshold. As will be seen later, the maximum field amplitude is the order of $10^{10}$V/m, which is similar to a previously reported example [28]. To be explicit the FDTD method here uses a typical silicate glass for which the maximum field is close to the optical damage threshold but there are other nonlinear materials and perhaps metamaterials that could support the solitons with greater ease. A final point to make in this connection is that the FDTD method is a time-development of the field evolutions and what is seen in the pictures below are snapshots. Hence, even if the fields are high, the FDTD method gives an impressive view of the optical situation at each stage of the evolution. On the basis of the arguments given above the numerical results do give a clear idea of how an optical beam interacts with a given surface.

The (X,Y) plane is split into an upper RHM dispersion-free space and a lower dispersive LHM space. To be specific, the data selected for the RHM takes into account the fact that it can become nonlinear and sustain a spatial soliton, in which self-focusing is opposed by diffraction. The spatial solitons are sourced in this type of nonlinear host by adopting the initial condition:

$$E_z(X=0,Y) = \xi \frac{1}{kw} \frac{1}{\sqrt{n_0 n_2}} sech\left(\frac{Y-Y_0}{w}\right) \quad (9)$$

where $k$ is the wavenumber, $w$ the characteristic width, and $Y_0$ is the beam centre position. This equation illustrates a fundamental difficulty connected with the launched of narrow spatial solitons. At first sight it would appear that the parameter $\xi$ is unnecessary. Unfortunately without some scaling of this kind the narrowness of the beams creates such a diffraction-dominated scenario that self-focussing cannot make itself apparent [5]. It is not appropriate to go any further into this but the typical value of $\xi$ is 1.2. All the data were selected to be specific and fits in with the 'desire' of the FDTD method to use real coordinates. Similar outcomes can be generated with other data, so this choice is not prohibitive in any way. The ultra-narrow beam widths used here are of approximately one wavelength ($\lambda$). This value is sufficient to achieve some level of stability for the beams, as indicated earlier, and prevents them from blowing up under the duress of numerical noise, before they can be usefully deployed, Figure 2 shows some of the properties that can be expected from the use of such narrow beams. It is clear from the figure beams of the order of a wavelength in width propagate over a sufficient number of cells to be of great value for interaction purposes. In the same figure it is demonstrated that beams of the order of half a wavelength are prone to break up at less than half of this distance. Pair wise interactions have an even more destructive effect upon stability.

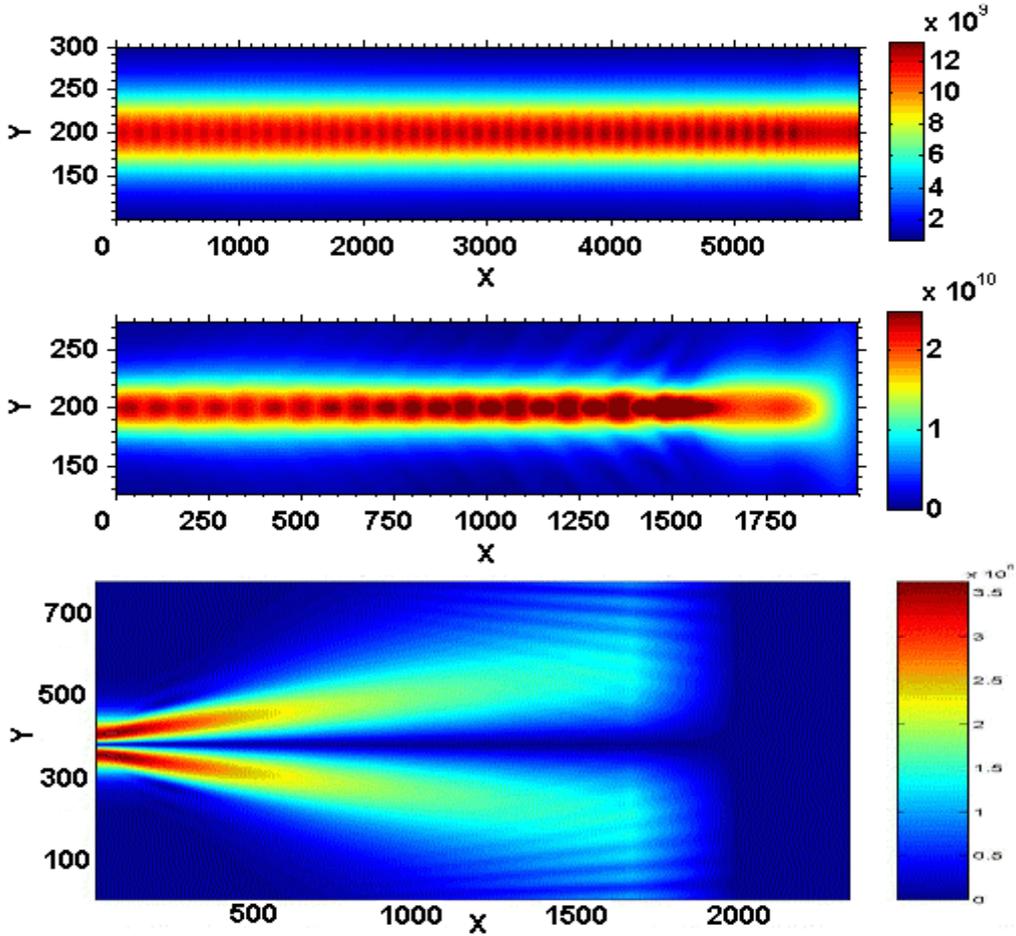

**Figure 2:** The upper figure demonstrates stable propagation of a 1 $\lambda$ wide beam soliton over a substantial distance compared to the distance shown in the middle figure for a beam width of 0.5 $\lambda$. The bottom panel indicates the destructive effects caused by interactions. $x$ and $y$ are measured in units of cell $\lambda/33$.

If different angles of incidences to an interface are required then an additional spatial phase factor is used. In this case solitons are *injected* into the grid at $x = 0$ with a sinusoidal temporal profile. An interesting and further FDTD advantage is now apparent. It is possible to have two temporal functions: one to describe the expected oscillatory fields and another to simulate a 'switch-on' period where the beam amplitude ramps up to the desired value. By varying the 'switching' duration, the initial beam frequency bandwidth can be controlled. Given the inherent time-domain nature of the FDTD, this allows an additional avenue of exploration, in that time- dependent effects can be explored through their dependence on the spectral content they are circumscribed by.

## 4. NUMERICAL INVESTIGATIONS OF SURFACE BEHAVIOUR

The generation of linear surface plasmon-polaritons has been extensively addressed over the last three decades [1]. Basically, there will be a problem surrounding the generation of surface waves by directing an excitation beam of electromagnetic energy onto a surface without taking care of the necessary phase-matching conditions. For example, attempts to launch surface waves onto a metal surface from air or a vacuum fail to provide the correct momentum for their generation without phase-matching assistance. The latter can be provided by a prism [2], or thin film arrangement, or some kind of grating or end-fire coupling. As stated earlier, for an entirely RHM set of materials, the surface waves are TM-polarised. Furthermore, the relative permittivities of the interfacing materials must possess opposite signs. As seen in section 2, however, a LHM surface can also support plasmonic surface waves but this time they can be TE *or* TM modes. At the boundary, there will be a discontinuity of the slope of the $E_z$ electric field component in one case and of the $E_x$ component in the other case. Both should be sensitive to the nonlinearity that is supporting an incident spatial solitons that happens to be arriving through a non-dispersive RHM. This sensitivity is almost certain to influence the surface wave power flow direction during the interaction time of the excitation beam with the surface. It can even be determined by the instantaneous value of $\varepsilon_c$ that is given by Eq.(4).

The simulations, use *TE-polarised bright spatial solitons* to interrogate interfaces between a RHM and a LHM and all are designed to address the optical domain of frequencies, even though a rather lot has appeared in the literature about the GHz range. The decision to address the optical domain is driven partly by the current optical evidence of left-handedness, and partly because other applications that rely upon the 'focusing of light using negative refraction' [sic][25] are evident in the literature. The quest to find suitable left-handed materials suggests that composite materials based upon nanowires [23, 24] may well turn out to be very successful.

The first simulation is shown in Fig.3 and mimics the famous Otto configuration [2]. It has an air gap separating the LHM from the RHM. The generation of a surface wave involves phase matching the wave number of the surface wave that can be supported by the LHM to the wave number component of the incoming beam that is parallel to the interface. The procedure is as follows. The frequency is selected to be the value cited earlier in the text and this gives the surface mode wave number $k = \frac{\omega}{c}\sin\theta$, where $\theta$ is the angle of incidence of the incoming beam and $c$ is the velocity of light in the medium. Given $\omega$ and $k$, a value of $\mu$ can be selected that then forces a value for $\varepsilon$ obtainable using equation (5). The values of $\mu$ and $\varepsilon$ can then be adjusted to give forward or backward waves, with the selection criterion being centred upon $\varepsilon_c$.

The thickness of this gap is 10 cells in the FDTD space used and it is sandwiched between a typical nonlinear RHM silicate glass and a linear LHM in which the resonant frequencies are tuned to give, for the assumed carrier frequency $\omega_c = 1.77 \times 10^{15}$ s$^{-1}$, $\mu = -0.29$ and $\varepsilon = -5.81$. These values imply that $\varepsilon_c$ = 9.56. Hence, since the air gap is RHM with $\varepsilon_b$ = 1 then $\varepsilon_b < \varepsilon_c$ and *linear backward surface waves are expected*. This is because the air/LHM is a linear /linear interface now. This is precisely what is observed in Fig.3a, in which the beam is incident at an angle of 30º. Not only does this angle place the calculation well beyond any paraxial approximation, thus emphasising the power of the FDTD method, but the generation of the backward wave is a justification of the simple theory outlined in section 2. The narrow soliton overcomes diffraction and this is clearly demonstrated.

The surface of the LHM can also support a forward surface wave. In fact, the data set $\mu$ = -1.2800 and $\varepsilon$ = -0.6963 leads to $\varepsilon_c$ = 0.48 and therefore $\varepsilon_b > \varepsilon_c$. In this case *linear forward surface waves are expected*. This is observed in Fig.3b. For completeness, Fig.3c shows what to expect if a spatial soliton is not used. Massive diffraction is observed that is exaggerated by the beam narrowness and the Rayleigh length being the order of 288 cells on the FDTD grid. The issue here is not about the amplitude of the excited wave but about the integrity of the reflected beam. As can be seen, even though the amplitude of the excited wave is similar in both the linear and nonlinear case the integrity, in terms of damage due to diffraction, of the reflected beam is markedly different

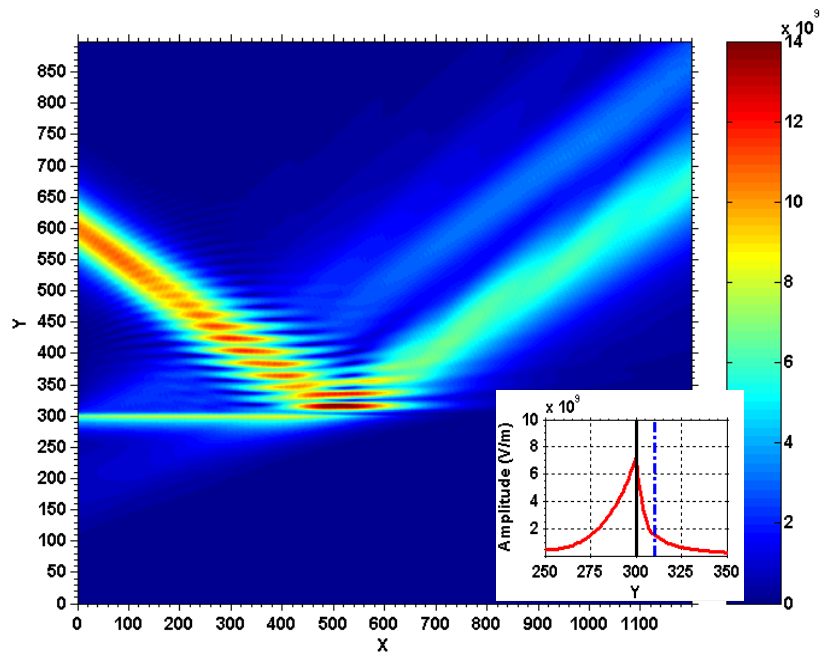

**Figure 3a:** Time-averaged electric field amplitude over the time step $T$ = 12000 units. LHM/air interface at $Y$ = 300, air/ nonlinear glass interface at $Y$ = 310. Inset displays the characteristic surface wave profile taken at $X$ = 0 with interfaces marked by solid (LHM/air) and dashed (air/glass) lines. Data: $\mu$ = -0.29 and $\varepsilon$ = -5.81. $x$ and $y$ are measured in units of cell $\lambda$/33.

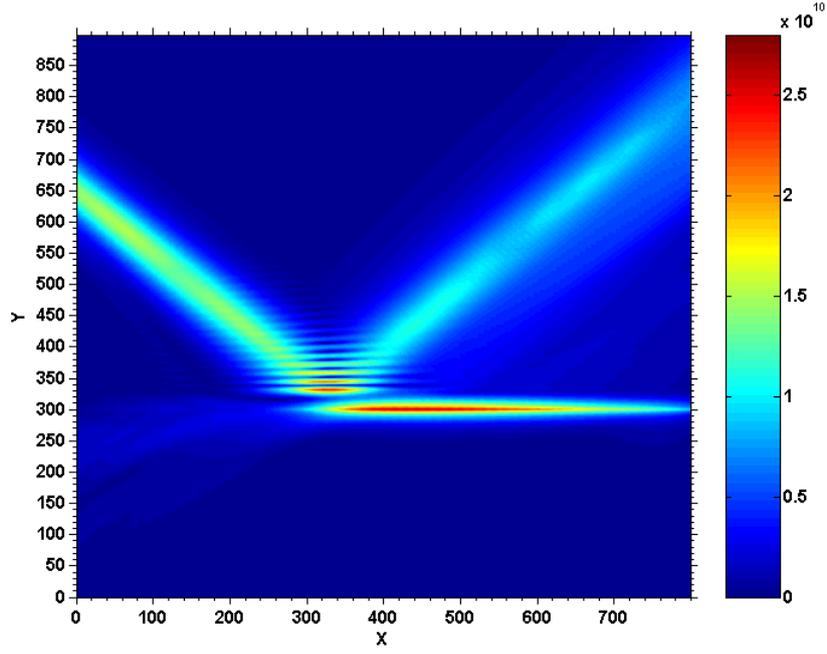

**Figure 3b:** Time-averaged electric field amplitude over the time step $T = 7000$ units. LHM/air interface at $Y = 300$, air/linear glass ($n = 1.53$) interface at $Y = 330$
Data: $\mu = -1.2800$ and $\varepsilon = -0.6963$, 45º beam angle with air gap of 30 spatial cells. $x$ and $y$ are measured in units of cell $\lambda/33$.

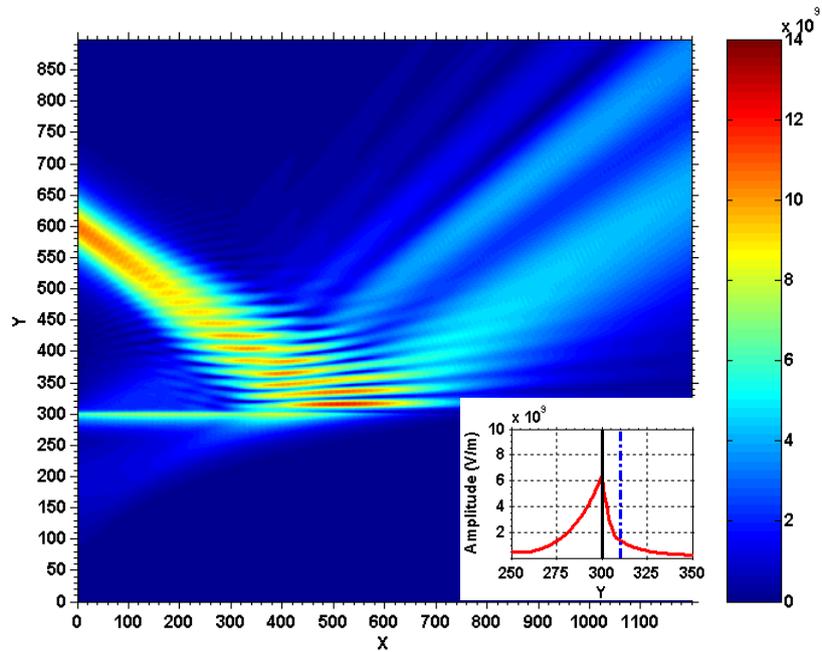

**Figure 3c:** Time-averaged electric field amplitude over the time step $T = 12000$. LHM/air interface at $Y = 300$, air/linear glass ($n = 1.53$) interface at $Y = 310$,

Inset displays the characteristic surface wave profile taken at $X = 0$ with interfaces marked by solid (LHM/air) and dashed (air/glass) lines. Data: $\mu = -0.29$ and $\varepsilon = -5.81$. $x$ and $y$ are measured in units of cell $\lambda/33$.

It is interesting that in Figs. 3a and 3c, regardless of whether the RHM is nonlinear or not, a second 'shadowy' beam is just visible above and to the left of the primary reflection. This second beam appears to be caused by a surface wave-enhanced negative Goos-Hänchen shift, of the kind predicted earlier [45]. Basically because the beam can be expanded into a set of plane waves then the question of what the generation of surface modes will do to enhance or diminish the linear Goos-Hänchen shift that could be expected from a beam incident upon a dispersionless medium should be addressed. The answer [45] is that large shifts can be expected if surface waves are generated by the beam. This large, or giant, shift can also be predicted for the LHM [22]. Although this issue needs a fuller numerical investigation and will be the topic of another publication, it can be seen in Fig, 4 that driving the surface wave closer to resonance results in a dramatic increase in the surface wave amplitude and a dominance of the 'shift-enhanced' second beam. The figure shows the location of the expected position of the beam.

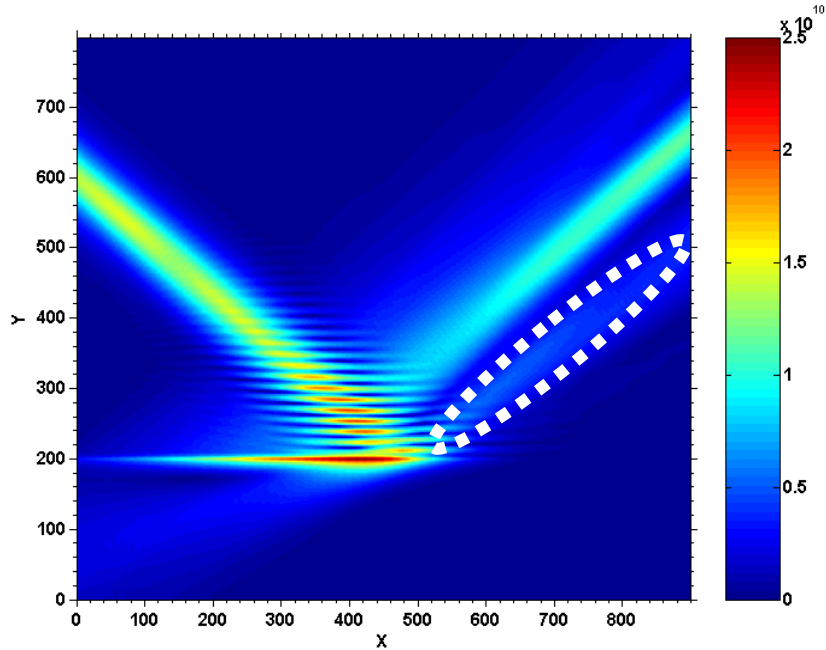

**Figure 4:** Time-averaged electric field amplitude over the time step $T = 10000$.
LHM/air interface at $Y = 300$, air/linear glass ($n = 1.53$) interface at $Y = 310$,
Dashed curve indicates expected position of beam in non resonant case.
Data: $\mu = -0.5000$ and $\varepsilon = -2.5605$, beam angle to 40°. $x$ and $y$ are measured in units of cell $\lambda/33$.

Removing the air gap prevents the launching of a propagating surface wave. However, the interface becomes nonlinear whenever the nonlinear medium supporting the spatial solitons sits directly upon the LHM. It is interesting then that a localised excitation is still formed in the LHM and should have the signature of a nonlinear surface wave. If the data is critically selected, the deposited energy in the LHM should possess some of the characteristics shown in Figs. 3. The behaviour of the beam can be made to demonstrate a sensitive dependence upon the ratio of the power levels in the respective nonlinear and the left- handed media. Indeed, the intensity delivered by the spatial solitons directly to the LHM surface results in a surface deposition of energy that causes the effective $\varepsilon_b$ to oscillate from its linear value. The power in the arriving beam causes a dynamic readjustment of the refractive index near to the surface. In fact the linear index in the upper medium can be increased by the power and as time progresses it can drop back temporarily towards its linear value. This is what is meant by the oscillation and at no time can $\varepsilon_b$ fall below its linear value. For the data used to generate Figs.5, 6 $\varepsilon_c$ = 2.25, so with the linear part of the permittivity in the medium supporting the solitons being $\varepsilon_b$ = 2.34 it is expected that the energy redistribution taking place in the vicinity of the surface will cause $\varepsilon_b$ to increase further above $\varepsilon_c$ and then relax back towards the original $\varepsilon_b$. In which case, the quasi-nonlinear-surface wave that is making its appearance beneath the contact area will appear to oscillate back and forth with time i.e. a kind of '*dancing nonlinear surface wave*' will be created. In fact, what is happening is that an attempt is being made to create surface excitations with different speeds and that the latter are determined by the extent of the difference between $\varepsilon_b$ and $\varepsilon_c$. This can be appreciated from an inspection of Figs. 5, 6 and 7. At the point where the beam gets compressed by the interface, the intensity changes, this causes the localised energy to change its position and there is also the possibility of new frequency generation. The 'dancing' activity takes place over about 300 cells, equating to approximately 10μm, or one Rayleigh length, so that the energy *dancing* occurs over, a relatively considerable distance, and initially takes place over a time the order of 400fs but will continue to 'oscillate'.

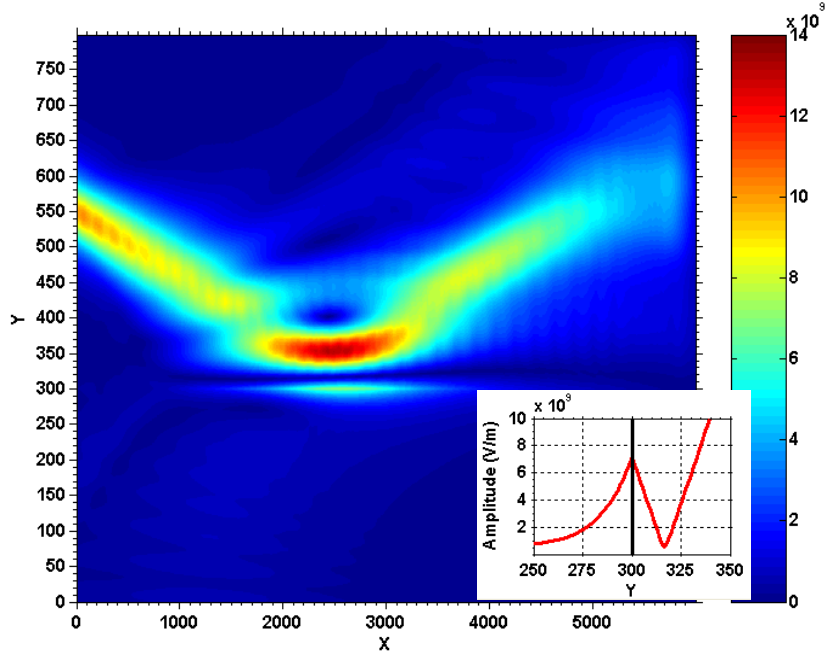

**Figure 5:** Time-averaged electric field amplitude over the time step $T = 19000$. glass/LHM interface at $Y = 300$. Inset displays the characteristic surface wave profile taken at $X = 2600$ with the interface marked by a solid line. Data: $\mu = -1$ and $\varepsilon = -2.25$ beam angle 5º. $x$ and $y$ are measured in units of cell $\lambda/33$.

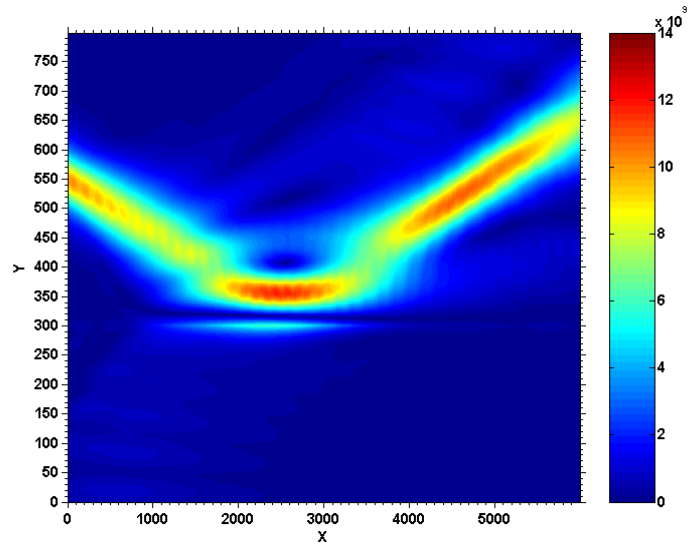

**Figure 6:** Time-averaged electric field amplitude over the time step $T = 25000$. glass/*LHM* interface at $Y = 300$. Data: $\mu = -1$ and $\varepsilon = -2.25$ beam angle 5º. $x$ and $y$ are measured in units of cell $\lambda/33$.

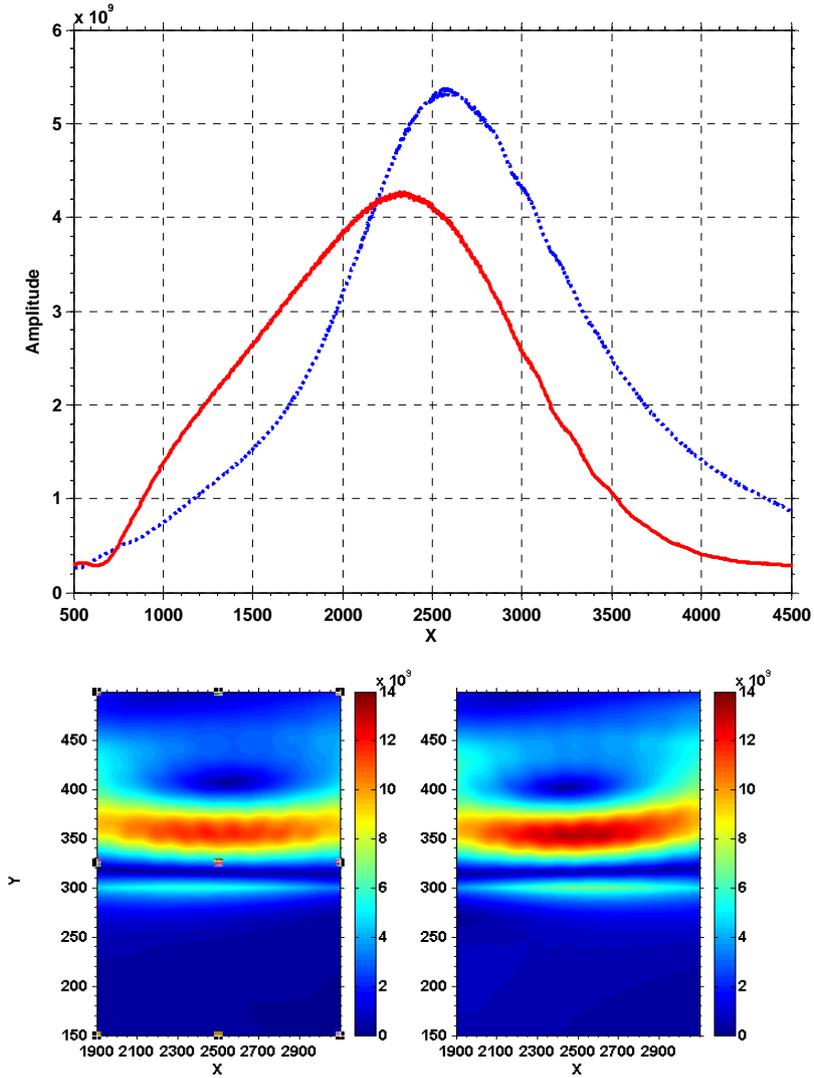

**Figure 7:** a) Profiles extracted along glass/LHM interface at $Y = 300$ from time-averaged electric field amplitudes for time-step $T = 19000$ (dotted) and $T = 25000$ (solid). b) Zoomed time- average of the electric field for the time- steps 25000 and 19000. Data: $\mu = -1$ and $\varepsilon = -2.25$ beam angle 5º. $x$ and $y$ are measured in units of cell $\lambda/33$.

Since these effects are as a result of temporal dependence it is wise to utilise the availability of beam 'switch on' provided by the FDTD scheme to explore the effects of frequency content i.e. bandwidth. All of the results considered so far are for a beam that is 'unrealistically' switched on over a zero number of periods i.e. the amplitude of field oscillation is instantaneously a maximum at time $T = 0$. Using a switch-on mechanism in which the amplitude of oscillation is ramped up to its maximum value over 30 periods may be more desirable. 30 periods has been suggested as a value that provides a way of reducing the bandwidth of the system sufficiently for the LHM to reach its steady state as quickly as possible [32]. To investigate the truth of this matter, the foregoing simulations have been repeated with this level of switch-on time but it has no appreciable effect on the general behaviour of the system. Most of the energy is carried at the resonant frequency to which the LHM has been tuned. As such, the decrease in bandwidth has no observable effect. This is also consistent with the fact that the dispersion curves for the model used here are considerably flatter than those that will appear for the usual microwave model.

## Conclusions

This investigation uses very narrow spatial soliton beams to provide visual evidence, via a comprehensive FDTD set of numerical experiments, of how surface waves can be generated upon a left-handed interface. It is emphasised that narrow spatial solitons are excellent tools for probing surfaces. They can be the order of a wavelength in width, will be deployable in nonparaxial applications, maintain their shape against any onset of diffraction and have Rayleigh lengths that are small enough for the kind of sub-wavelength all-optical chip applications of the future. At these beam widths, perturbations can destroy them, even in 1D, but this has been addressed here and TE-polarised beams have been adopted. Numerically induced errors lead to a maximum propagation length for such beams, in a manner similar to unavoidable material fluctuations in a 'real world' practical sense. In this way, the FDTD mirrors experimentation in outlining some of the restrictions that a fully theoretical approach sometimes may not consider.

The FDTD computer simulations, for the Otto, prism-air-gap, configuration, beautifully demonstrate the way that *backward or forward linear surface waves* can be generated on a left-handed surface, in agreement with a theoretical prediction. These are computer simulations using FDTD to launch surface waves, using TE solitons in the Otto manner. They show the generation of forward and backward surface waves. With the air gap removed, the beams do not generate a strict surface wave but they do deposit energy into the left-handed medium, which have properties that are the signature of *nonlinear surface waves*. This is because now a nonlinear/linear interface is created. The ability of the nonlinearity to change the dielectric permittivity in a fluctuating manner results in a change in the velocity and hence gives a surface excitation that moves slowly or quickly. Qualitatively, this gives the impression of what has been referred to as a 'dancing' excitation. The FDTD method has an inherent strong capability to model advanced or potential engineering problems and complex geometries will be the subject of a future publication.

## Acknowledgments

We would like to thanks to Dr Ming Xie for her kind support in letting us use her workstation to perform some of the simulations shown here. In addition, the authors wish to thank one of the referees for drawing our attention to a paper submitted in the same month as this paper. It is I.V. Shadrivov, R.W. Ziolkowski, A.A. Zharov and Y.S. Kivshar, "Excitation of guided waves in layered structures with negative refraction," Opt. Exp. **13(2)**, 481-492 (2005).